\documentclass[
 article,
 twocolumn,
 groupedaddress,
 showpacs,
 preprintnumbers,
 longbibliography,
 amsmath,
 amsthm,
 amssymb,
 aps,
 prab,
 floatfix,
]{revtex4-1}

\usepackage{graphicx}					
\usepackage{verbatim}					
\usepackage{dcolumn}                    
\usepackage[many]{tcolorbox}
\usepackage[ruled,vlined]{algorithm2e}
\usepackage{color}
\usepackage{fontawesome}
\usepackage{adjustbox}

\usepackage{orcidlink}                     

\newcommand{\ds}{\displaystyle}             

\newcommand{\dd}{\mathrm{d}}                
\newcommand{\pd}{\partial}                  

\newcommand{\Rs}{\mathcal{R}}                 

\newcommand{\const}{\mathrm{const}}
\newcommand{\ob}{\mathcal{O}}                

\newcommand{\K}{\mathcal{K}}                

\begin{document}

\title{
Geometry of Almost-Conserved Quantities in Symplectic Maps:
\\Approximate Invariants in Nonlinear Accelerator Systems
}
\author{T.~Zolkin\,\orcidlink{0000-0002-2274-396X}}
\email{iguanodyn@gmail.com}
\affiliation{Independent Researcher, Atlanta, GA 30341}
\author{S.~Nagaitsev\,\orcidlink{0000-0001-6088-4854}}
\email{snagaitsev@bnl.gov}
\affiliation{Brookhaven National Laboratory, Upton, NY 11973}
\affiliation{Old Dominion University, Norfolk, VA 23529}
\author{I.~Morozov\,\orcidlink{0000-0002-1821-7051}}
\email{i.morozov@corp.nstu.ru}
\thanks{currently at Elettra Sincrotrone Trieste}
\affiliation{Novosibirsk State Technical University, Novosibirsk 630073, Russia}
\author{S.~Kladov\,\orcidlink{0000-0002-8005-9373}}
\email{kladov@uchicago.edu}
\affiliation{University of Chicago, Chicago, IL 60637}

\date{\today}

\begin{abstract}
We present a perturbative method for constructing approximate
invariants of motion directly from the equations of discrete-time
symplectic systems.
This framework offers a natural nonlinear extension of the classic
Courant-Snyder (CS) theory for systems with one degree of freedom
--- a foundational cornerstone in accelerator physics now spanning
seven decades and historically focused on linear phenomena.
The original CS formalism emerged under conditions where
nonlinearities were weak, design goals favored linear motion,
and analytical tools --- such as the Kolmogorov-Arnold-Moser (KAM)
theory --- had not yet been fully developed.
While various normal form methods have been proposed to treat
near-integrable dynamics, the approach introduced here stands
out for its conceptual transparency, minimal computational
overhead, and direct applicability to realistic systems.
We demonstrate its power and versatility by applying it to
several operational accelerator configurations at the Fermi
National Accelerator Laboratory (FermiLab), illustrating how
the method enables fast, interpretable diagnostics of nonlinear
behavior across a broad range of machine conditions.
\end{abstract}

\maketitle

\section{Introduction}

Since its inception, the foundational theory of motion in circular
particle accelerators has been built around linear dynamics.
The celebrated Courant-Snyder (CS) theory~\cite{courant1958theory},
developed in the 1950s, provided a powerful analytical framework
for understanding phase space evolution in systems weakly
perturbed from harmonic oscillations,
forming the theoretical backbone of modern accelerator design
\cite{SYLee4,wiedemann2015particle}.
However, this linear formalism is increasingly strained in the
face of modern challenges: higher beam intensities, smaller beam
emittance, tighter stability margins, and the pervasive presence
of nonlinear magnetic elements in advanced accelerator facilities
such as the Integrable Optics Test Accelerator
(IOTA)~\cite{DanilovNagaitsev,jarvis2022experimental},
as well as in proposed next-generation machines like
the Muon Collider~\cite{PhysRevLett.134.160001} or
the FCC \cite{fcc1,fcc2,fcc3}.

Despite the critical role of nonlinear effects in determining
long-term stability and beam quality, a general analytical
extension of the Courant-Snyder formalism to strongly nonlinear
regimes has remained elusive, even for one degree of freedom.
In particular, the twist expansion of action-angle variables
achieved via Lie-algebraic methods is compatible with numerous
invariants of motion, all of which are conserved to the same
order of approximation (see~\cite{zolkin2026geometry} for
details).
In the widely used formalism of Birkhoff normal forms, this
ambiguity is commonly addressed by isolating resonances, which
requires the manual selection of a specific normal form.
In this work, we present an approach that extends this
understanding:
a systematic perturbative framework for constructing approximate
invariants in symplectic, reversible maps~\cite{roberts1992revers},
in which a different averaging strategy allows multiple resonances
to be retained up to a given perturbative order.
We note that:
(i) direct comparisons of our method with resonant normal forms
(isolated-resonance approximation) can be made by juxtaposing
our Fig.~20 in~\cite{zolkin2026geometry} with Fig.~6
of~\cite{BAZZANI1993};
and (ii) our averaging technique recovers exact invariants in
integrable systems, providing a natural benchmark for its
applicability and demonstrating consistency with established
results.
Prominently, the method is compact both in formulation and
implementation, relying on a single guiding principle:
approximate invariance.

\newpage
Consider a symplectic transformation
$\mathrm{T}:\mathbb{R}^2\mapsto\mathbb{R}^2$
describing a dynamical system with one degree of freedom,
characterized by canonical coordinate $q$ and conjugate
momentum $p$.
In the integrable case, an exact invariant $\K[p,q]$ exists
and satisfies
\[
    \K[\mathrm{T}(p,q)] - \K[p,q] = 0,
    \qquad\qquad \forall (q,p)\in\mathbb{R}^2.
\]
We generalize this to non-integrable scenario by relaxing the
right-hand side to allow deviations of order $\ob(\epsilon^{n+1})$,
where $n$ defines the order of perturbation.
From this single condition, we construct a parameterized function
that captures the essential geometry of phase space --- revealing
the topology of nonlinear motion, the structure of island chains,
separatrices, and the amplitude dependence of the rotation number.
As discussed in the first part of this series~\cite{zolkin2026geometry}, the resulting
approximate invariant also preserves two families of reversibility
induced symmetries up to the same order, reinforcing both its
geometric fidelity and consistency.

The resulting method is both powerful and remarkably efficient,
suitable for both numerical and analytical computations.
All examples presented in this work --- spanning multiple
accelerators from the FermiLab facility --- are generated
from a single, closed-form function parameterized by the elements
of the accelerator lattice.
More precisely, it depends only on the coefficients of the
discrete-time transformation (the {\it one-turn map}) that
describes the evolution of a particle over a single revolution
with respect to a reference trajectory.
This function acts as a nonlinear generalization of the
Courant-Snyder invariant, enabling rapid visualization of stability
boundaries, identification of resonant structures, and quantitative
prediction of long-term dynamics --- without requiring particle
tracking or extensive numerical simulation.
When combined with the Danilov's theorem
\cite{nagaitsev2020betatron,nagaitsev2020betatronER}, this technique
can be connected to standard methods, allowing the construction
of a set of approximate action-angle variables and
nonlinear betatron tune.

In the first part of this series~\cite{zolkin2026geometry}, we
developed the theoretical foundations of our perturbative approach
by introducing the central concept of an approximate invariant
constructed order by order in the small parameter $\epsilon$.
This parameter naturally emerges from the smallness of oscillation
amplitudes and does not, in itself, require the nonlinear forces
to be perturbatively weak.
We demonstrated that these invariants respect the discrete
reversibility symmetries --- serving as a discrete analogue of
Noether's theorem --- and benchmarked the results against standard
Lie-algebraic normal-form methods near a fixed
point~\cite{MichelottiLeo1995,Forest1998,Berz1999}.
We also showed that, at each perturbative order, the approximate
invariant is not unique but is determined only up to arbitrary
terms composed of powers of the zeroth-order (linear)
Courant-Snyder invariant (see~\cite{zolkin2026geometry} for
details).
Regardless of the choice of these coefficients, the twist
expansion obtained from the standard Lie normal-form procedure
remains satisfied.
To resolve this arbitrariness, we introduced an averaging
procedure that minimizes the fluctuations of the approximate
invariant.
This not only fixes the previously undetermined coefficients but
also effectively resolves degeneracies associated with rational
rotation numbers, making the theory applicable to singular (with
diverging twist coefficients) and to $n$-island chain resonances.
We note that this averaging differs from the traditional approach
used for isolated resonances~\cite{BAZZANI1993} and can properly
account for all resonances up to a given order of approximation,
thereby reinforcing both the geometric and physical foundation of
the method.

The second part of~\cite{zolkin2026geometry} extends the
discussion from local, infinitesimal dynamics to large-amplitude
motion, aiming for a global reconstruction of (quasi-) integrable
structures in both integrable and chaotic systems.
Applying the method to several representative maps --- 
including the
McMillan, Suris exponential, and H\'enon quadratic and cubic area-preserving transformations
--- we demonstrated both convergence and resonance compatibility.
The averaged invariant not only exhibited the expected asymptotic
behavior near the fixed point but also reproduced exact global
results for integrable systems, successfully capturing
separatrices and other invariant structures beyond
(see~\cite{zolkin2026geometry} for details).
A detailed comparison with the Square Matrix
method~\cite{hua2017square,square}, which is equivalent to the
full normalization in the language on normal forms, showed
comparable local accuracy but superior global stability and
resonance performance in the present approach.

In this final part, we briefly restate the method for general
planar maps and then move directly to demonstrating its practical
power.
We apply the framework to realistic dynamical models drawn from
accelerator lattice configurations across the FermiLab complex.
Specifically, we examine three major facilities:
the Integrable Optics Test Accelerator (IOTA),
the Main Injector (MI), and the Mu2e Delivery Ring.
These systems exhibit a wide range of nonlinear behavior,
including high-order resonances and large-amplitude instabilities.
In each case, our method yields rapid and interpretable diagnostics
that align closely with results from full-scale particle tracking
--- while dramatically reducing computational demands.

Beyond its immediate practical utility, this approach offers a
scalable framework for nonlinear design and optimization ---
potentially transforming how we diagnose, predict, and ultimately
control stability in high-performance machines.
It introduces a new paradigm for describing nonlinearity.
Just as the linear CS theory allows one to bypass the detailed
specification of all quadrupole magnets --- often numbering in
the hundreds --- by reducing the system to a small set of Twiss
parameters that describe the one-turn map and connect directly
to key dynamical quantities such as betatron tune and beam
envelope, our nonlinear extension leverages the cumulative effect
of all nonlinear elements, condensed into the coefficients of the
one-turn map, to construct approximate invariants.
Thereby enabling a compact and highly informative representation
of the system's dynamics.

\newpage
\section{Perturbation method}

Consider a map of the plane $\mathrm{T}$:
$\mathbb{R}^2\mapsto\mathbb{R}^2$, which is symplectic
(i.e., it preserves area and orientation) and admits a
power series expansion:
\[
\begin{array}{cc}
\ds q' = A_{1,0}\,q + A_{0,1}\,p + A_{2,0}\,q^2 + A_{1,1}\,q\,p +
A_{0,2}\,p^2 + \ldots,\\[0.25cm] 
\ds p' = B_{1,0}\,q + B_{0,1}\,p + B_{2,0}\,q^2 + B_{1,1}\,q\,p +
B_{0,2}\,p^2 + \ldots.
\end{array}
\]

\noindent$\bullet$
We begin by introducing a general polynomial ansatz for the
approximate invariant of order $n$:
\[
\K^{(n)}[p,q] = \K_0 + \epsilon\,\K_1 + \ldots + \epsilon^n\,\K_n,
\]
where $\K_m$ is a homogeneous polynomial of degree $(m+2)$:
\[
\K_m[p,q] = \sum_{\substack{i,j \geq 0 \\ i + j = m+2}}
C_{i,j}\,p^i q^j.
\]
To determine the coefficients $C_{i,j}$, we require that the
approximate invariant be conserved up to $\ob(\epsilon^{n+1})$:
\[
\Rs_n = \K^{(n)}[p',q'] - \K^{(n)}[p,q] = 
    \overline{\Rs_n}\,\epsilon^{n+1} +
    \ob(\epsilon^{n+2}).
\]
The lowest-order contribution, which can also be derived via
standard linearization, is defined only up to a constant
multiplier
\begin{equation}
\label{math:K0}    
\K_0[p,q]=C_0\left[
    A_{0,1}\,p^2+(A_{1,0}-B_{0,1})\,p\,q-B_{1,0}\,q^2
\right].
\end{equation}
The {\it seed} coefficient $C_0$ can be set to unity or chosen
specifically to eliminate resonant denominators.

\noindent$\bullet$
Due to under-determinacy, the invariant (after all two-indexed
coefficients $C_{i,j}$ have been fixed) is known only up to a
series:
\[
    C_0\,\K_0 + C_1\,\K_0^2 + \ldots.
\]
Nevertheless, the approximate invariance condition holds to the
required order, and the twist coefficient $\tau_k$ converges at
order $n = 2\,k + 2$.

\noindent$\bullet$
The remaining coefficients $C_k$ (for $k>0$) are obtained via the
{\it averaging procedure}, which minimizes the residual error.
After transforming to the eigenbasis of the Jacobian,
$(q,p)\rightarrow(Q,P)$, the linear part of the map becomes a pure
rotation, and the zeroth-order invariant takes the form
\[
\K_0[P,Q] = P^2 + Q^2.
\]
The coefficients $C_k$ are then determined by solving another
system of linear equations
\[
    \frac{\dd}{\dd C_k}\,I_n = 0,
\]
where the integral
\[
I_n = \int_0^{2\pi} \overline{\Rs_n}^2[\rho,\psi] \,\dd\psi.
\]
Here, $(\rho,\psi)$ are polar phase space coordinates defined by
$Q = \rho\,\cos\psi$ and $P = \rho\,\sin\psi$.

\subsection{Nonlinear Courant-Snyder theory}

In accelerator physics, the linear theory of motion uses a
parameterization in which the linearized one-turn map (the
discrete transformation that evolves a particle's phase space
coordinates over one full revolution)
\[
\begin{array}{cc}
\ds q' = A_{1,0}\,q + A_{0,1}\,p,\\[0.25cm] 
\ds p' = B_{1,0}\,q + B_{0,1}\,p,
\end{array}
\]
is written as
\[
\begin{bmatrix}
    q' \\ p'
\end{bmatrix} =
\begin{bmatrix}
\cos\Phi + \alpha\sin\Phi & \beta\,\sin\Phi \\
-\gamma\sin\Phi & \cos\Phi - \alpha\sin\Phi
\end{bmatrix}
\begin{bmatrix}
    q \\ p
\end{bmatrix}.
\]
The coefficients $\alpha$, $\beta$ and $\gamma$, collectively
known as the {\it Courant-Snyder/Twiss parameters}, are
not all independent: the symplectic condition imposes the
constraint $\beta\,\gamma = 1 + \alpha^2$;
conventionally, the {\it beta function} is chosen to be positive, 
$\beta > 0$.
$\Phi = 2\,\pi\,\nu_0$ is the {\it betatron phase advance}, with
$\nu_0$ being the bare {\it betatron tune} --- the unperturbed
rotation number.

This parameterization enables fast, linearized estimates of several
experimentally important quantities.
First, (i) one obtains the betatron tune, corresponding to the
frequency of horizontal or vertical oscillations around the closed
reference orbit (or the synchrotron tune in the longitudinal case):
\[
    \nu_0 = \frac{1}{2\,\pi}\,\arccos\frac{A_{1,0}+B_{0,1}}{2}.
\]
Second, (ii) one obtains the {\it Courant-Snyder invariant}
\[
    \K^{(0)}[p,q] = \gamma\,q^2 + 2\,\alpha\,q\,p + \beta\,p^2
\]
which provides a convenient parameterization of trajectories in the
phase space coordinates at a fixed location in the accelerator.
By propagating the Twiss parameters along the reference orbit,
this invariant allows the reconstruction of phase space
trajectories not only at each discrete turn but also continuously
along the machine's circumference, $s$.

This invariant can also be obtained using Eq.~(\ref{math:K0})
by selecting $C_0 = \beta^*/A_{0,1}$, where $\beta^*=\beta(s^*)$
is the beta function at a reference location, $s^*$.
Moreover, (iii) this scaling of the invariant,
\[
    \K^{(0)}[p,q] = C_0\,\K_0 = \mathrm{em}
\]
provides a direct correspondence between the invariant and the
{\it beam emittance} $\mathrm{em}$, which quantifies the area
occupied by the beam in phase space.
This value plays a crucial role in estimating performance metrics
of colliders such as luminosity, and, machines
involving strong dissipative forces, including potential muon
collider~\cite{PhysRevLett.134.160001} and facilities employing
electron or optical stochastic cooling~\cite{jarvis2022experimental}.
The perturbative method described earlier naturally extends this
linear framework.

\begin{figure*}[th!]
    \centering
    \includegraphics[width=\linewidth]{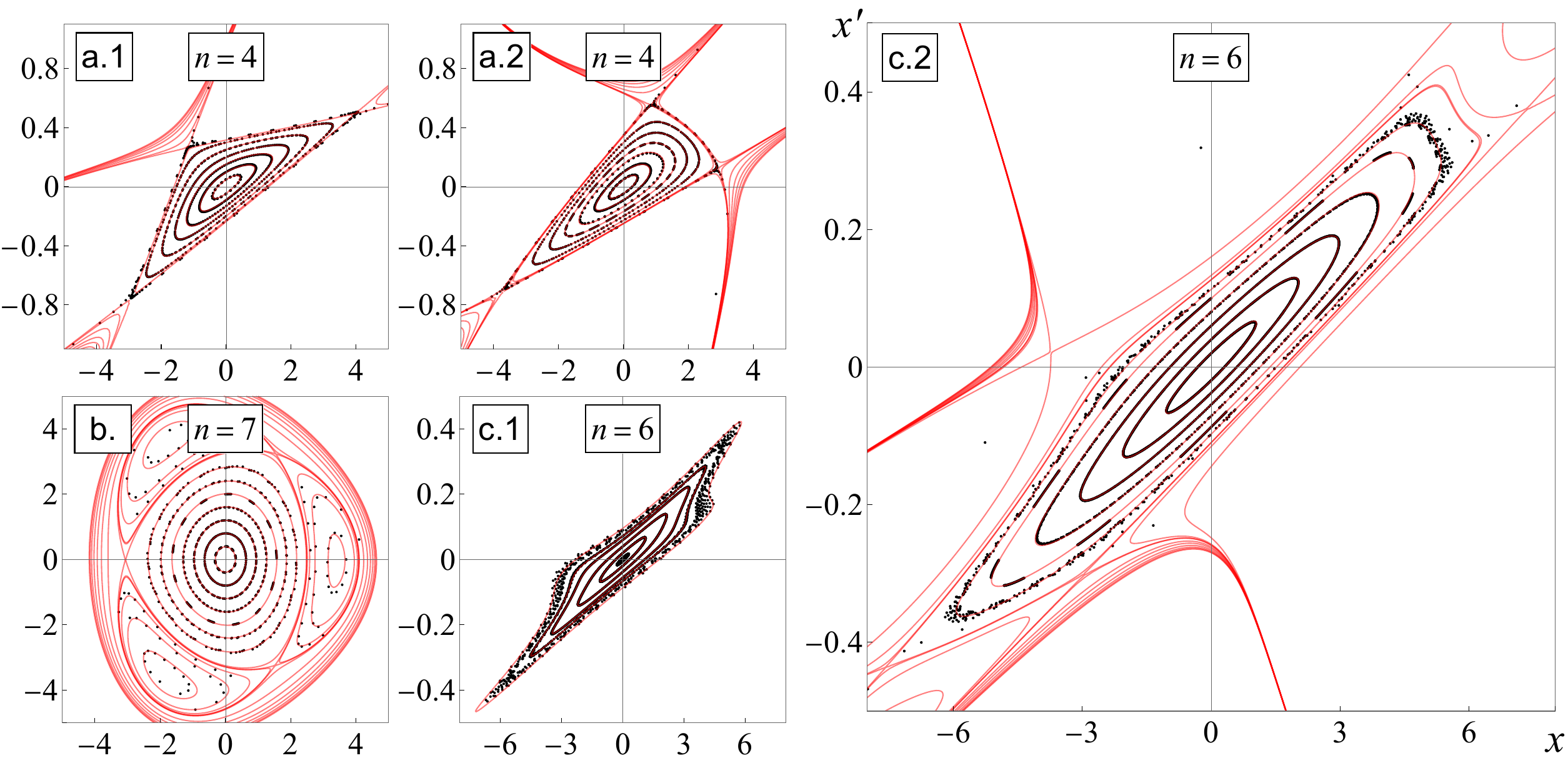}
    \caption{\label{fig:Accelerators}
    Comparison between numerically computed trajectories (black dots)
    and constant level sets of the approximate invariant (red curves)
    for various accelerator lattices at FermiLab.
    Panels (a.1,2) correspond to different configurations of
    slow extraction in the Delivery Ring for the Mu2e experiment,
    panel (b.) is for the IOTA ring, and panels (c.1,2) corresponds
    to the Main Injector ring.
    All plots display the horizontal phase space, with position $q$
    in millimeters and conjugate momentum $p = \dd q/\dd s$;
    for panel (b.), both $q$ and $p$ are normalized to the beam's rms
    size (in units of beam $\sigma$).
    }
\end{figure*}

\noindent
$\bullet$
The zeroth-order Courant-Snyder invariant is generalized to a
sequence of approximate invariants
\[
    \K^{(0)} \rightarrow \K^{(1)} \rightarrow \K^{(2)}
    \rightarrow \ldots.
\]
In many practical cases, low-order approximations suffice to
accurately describe particle motion, as seen in Fig.
\ref{fig:Accelerators}.

\noindent
$\bullet$
Similarly, the linear tune $\nu_0$ is replaced by its nonlinear
analogs:
\[
    \nu_0 \rightarrow \nu_1 \rightarrow \nu_2 \rightarrow \ldots.
\]
These are extracted from the approximate invariants using Danilov
theorem~\cite{nagaitsev2020betatron,nagaitsev2020betatronER}
\[
\nu_n = \int_q^{q'}
    \left(\frac{\pd \K^{(n)}}{\pd p}\right)^{-1}
\dd q \Bigg/
\oint
    \left(\frac{\pd \K^{(n)}}{\pd p}\right)^{-1}
\dd q
\]
or equivalently,
\[
\nu_n = \dd J_n'/\dd J_n,
\]
where $J_n$ and $J_n'$ are {\it action} and {\it partial action}
variables:
\[
J_n = \oint \frac{p\,\dd q}{2\,\pi},
\qquad\text{and}\qquad
J_n'= \int_q^{q'} \frac{p\,\dd q}{2\,\pi}.
\]
All integrals are taken along a constant level set
$\K^{(n)}[p,q] = \const$.
The integral $\int_{q}^{q'}$ spans one iteration of the map
(independent of the choice of $q$), while $\oint$ denotes a full
contour around a closed trajectory.

This series of approximations differs from the traditional
{\it twist expansion}:
\[
\nu(J) =
    \nu_0 + \tau_0 J +
    \frac{1}{2!}\,\tau_1 J^2 +
    \frac{1}{3!}\,\tau_2 J^3 +
    \ldots.
\]
At the same time, a comparison between the twist expansion and
the $n$-th order approximation series
\begin{equation}
\label{math:twist}
\nu_n(J) =
    \nu_0 + \tau_0^{(n)} J +
    \frac{1}{2!}\,\tau_1^{(n)} J^2 +
    \frac{1}{3!}\,\tau_2^{(n)} J^3 +
    \ldots.
\end{equation}
shows that for each even order $n=2\,m$, all coefficients up to
$\tau_{m-1}^{(n)}$ converge to their exact values $\tau_{m-1}$.

As $n$ increases, $\nu_n$ approaches the twist expansion $\nu(J)$,
Eq.~(\ref{math:twist}), and both approximations converge to the
integrable limit of the Twist map.
Yet, for non-integrable systems, this convergence comes at the cost
of a vanishingly small domain.
At the other extreme, when $n$ is small but nonzero, the two methods
differ in form but provide complementary insights into the system's
behavior.
The low-order {\it twist coefficients} like
\[
\tau_0 = \frac{\dd\nu}{\dd J},
\qquad\text{and}\qquad
\tau_1 = \frac{\dd^2\nu}{\dd J^2},
\]
capture key nonlinear features and are widely used in estimates of
collective beam stability~\cite{Gareyte1997,Shiltsev2017}.
At the same time, as demonstrated in the first part of this
manuscript~\cite{zolkin2026geometry} and in related studies
\cite{zolkin2024MCdynamics,zolkin2024MCdynamicsIII},
large-amplitude behavior and the edge of stability can be
significantly better captured using low-order nonlinear estimates
$\nu_1$ or $\nu_2$ (labeled as SX-1 and SX-2 respectively),
particularly near low-order singular (i.e., $\tau_0 = \infty$)
resonances such as integer, half-, third-, or quarter-integer.

\newpage
In the second part of~\cite{zolkin2026geometry}, we also
show that this method applies to the simply connected regions
bounded by resonance island chains appearing in non-singular
bifurcations.

\noindent
$\bullet$
Moreover, the approximate invariant can always be rescaled to
match the action, $\K^{(n)}[p,q] = J_n$, thus yielding a nonlinear
analog of emittance, just as in the linear case.

Before proceeding with examples, we note that the transformation
to the eigenbasis $(q,p)\rightarrow(Q,P)$ in Courant-Snyder theory
is achieved via the {\it Floquet transformation} and corresponding
{\it betatron amplitude matrix}
\[
\begin{bmatrix}
    q \\ p
\end{bmatrix} =
\mathrm{B}\,
\begin{bmatrix}
    Q \\ P
\end{bmatrix},          \qquad\qquad
\mathrm{B} =
\begin{bmatrix}
\sqrt{\beta} & 0 \\
-\alpha/\sqrt{\beta} & 1/\sqrt{\beta}
\end{bmatrix}.
\]
Once the coefficients $C_k$ are determined, they can be substituted
into the non-averaged invariant $\K^{(n)}[p,q]$, expressed directly
in physical coordinates.
This makes the formalism particularly convenient for experimental applications.
The approximate invariant with substituted coefficients $C_k$ is
referred to as the {\it averaged invariant} and denoted by
$\langle\K^{(n)}\rangle$.

Fig.~\ref{fig:Accelerators} illustrates this method for several
circular accelerators at Fermilab, which differ significantly in
both design and purpose: third-integer resonant slow extraction in
the Delivery Ring machine for the Mu2e experiment (panels a.1 and
a.2), the Integrable Test Optics Accelerator (IOTA) ring (b.), and
the Main Injector ring (c.1 and c.2).
In each case, we use the full machine map --- including all
nonlinear (thin or thick) elements --- derived from a lattice
file describing magnet locations and strengths, just as in each
standard accelerator simulation codes.

We compare direct numerical application of the map (black dots)
with the level sets of the averaged approximate invariant 
$\langle\K^{(n)}\rangle[p,q]$ (red curves).
Each red curve corresponds to the trajectory seeded by the matching
black point. If a red curve is not visible, it is fully overlaid by
the trajectory, emphasizing the excellent agreement.

In examples (a.1) and (a.2), dynamics are dominated by a singular
third-order resonance, and even a low-order approximation ($n=4$,
corresponding to $J^2$) gives outstanding agreement.
In the IOTA ring, we observe a non-singular third-order resonance
dominated by octupole-type nonlinearity, characterized by a
distinctive chain of islands.
In (c.), the Main Injector ring exhibits high-order resonances and
chaotic behavior, but even a sixth-order approximation
($n=6$, or $J^3$) accurately captures the stability region and
general structure of the phase space.

Finally, we stress that the examples shown rely on unrefined,
bare lattice configurations, which include imperfections and
design features not yet resolved or optimized experimentally.
These cases often prompt direct simulations --- aimed at improving
performance (as in a.1 and a.2) or reducing detrimental resonance
structures (as in c.).
That our method performs well in such cases underscores its value
as a robust and complementary tool for nonlinear dynamics analysis.

\section{\label{sec:Conclusion}Conclusions and possible generalizations}

While Fig.~\ref{fig:Accelerators} demonstrates the power and
versatility of our method through the applications to three
distinct accelerator rings at the FermiLab, it is important to
comment on its limitations and potential extensions.
First and foremost, this method is not intended to replace
traditional simulations during the final stages of machine design,
where collective effects must be accounted for.
Nevertheless, even in such regimes, it offers critical insight
into intrinsic single-particle dynamics --- such as
amplitude-dependent tune shifts --- which are essential for
estimating instability thresholds and determining the level of
Landau damping required for instability suppression
\cite{Shiltsev2017}.

A natural extension of this framework is to systems with more than
one degree of freedom, i.e., higher-dimensional symplectic maps.
While a treatment of nonlinear coupling is currently under
consideration, we note that even the theory of coupled linear motion
was historically developed significantly after the original
Courant-Snyder framework (see, e.g., coupled betatron theory~\cite{lebedev2010betatron} for two degrees of freedom and
\cite{glukhov2025} for three degrees).
Therefore, a few brief comments are in order.
In principle, the same perturbative approach can be applied;
however, for a system with two degrees of freedom, one must construct
two approximate invariants, each depending on four variables in a
four-dimensional phase space.
The trajectory then lies on a torus defined by the intersection of
two three-dimensional hypersurfaces, significantly complicating both
visualization and extraction of dynamical features.
Still, recent extensions of the Danilov theorem to higher dimensions
\cite{mitchell2021extracting} offer a promising framework to be used
in tandem with the construction of such invariants.

Despite these challenges, we have shown that even a two-dimensional
analysis often yields remarkable value.
This includes a range of practical scenarios:
(i) accelerators with flat beams, where the horizontal beam size
is much larger than the vertical;
(ii) systems dominated by a strong one-dimensional resonance, as
in the case of the Main Injector;
(iii) machines designed to exploit nonlinear instability for slow
resonant extraction, such as the Mu2e Delivery Ring; and
(iv) round beams, where this method can be directly applied to
radial degree of freedom, --- see~\cite{zolkin2024MCdynamicsII}
for further discussion.

The developed procedure is conceptually simple and grounded solely
in the first principle of approximate invariance.
In contrast to Lie-algebraic normal-form methods, it does not rely
on the exponential operator formalism and assumptions on the final
normalized system.
Instead, it provides a direct and geometrically transparent
construction of the invariant itself, order by order, making the
method physically intuitive and highly regular in application.

The framework is intrinsically linked to reversible symmetries:
the invariant is sought as a function that remains nearly unchanged
under two reversing symmetries up to a prescribed perturbative order.
In this sense, the method parallels a discrete Noether-like approach,
where reversibility substitutes for continuous symmetries.

While the commonly used full normalization
procedure~\cite{Dragt:Marylie,dragt1997lie,Forest1998,Berz1999Full}
primarily targets the computation of the amplitude dependent tune
shifts and resonance driving terms, the presented formulation
succeeds in both computation of the tune shifts by employing the
Danilov Theorem~\cite{nagaitsev2020betatron,nagaitsev2020betatronER} and attempting to
describe the stability boundary itself~\cite{zolkin2026geometry}.
Since the full normalization tries constructing conjugation to an
integrable system depending solely on the action variables, its
application to the problem of describing the stability boundary
is not well grounded.
Any resonant structure, like a chain on islands, inside the
stability domain constitutes a boundary that can't be passed and
leads to divergence of the complete normalization procedure.
The latter also implies the presented theory can produce a more
generic tune dependence picture that can cross the inner resonance
chains.
In attempting to describe the stability boundary the isolated
resonance model can be employed~\cite{BAZZANI1993}.
This requires careful selection of a resonance being studied and
a procedure to switch between resonances following parameters
change.
The map based perturbation theory is free of such limitations and
doesn't require an a priori selection of some resonance producing
a coherent expression for the invariant valid across the full
range of the intrinsic parameters.

In conclusion, by distilling the cumulative effect of all nonlinear
lattice elements into a single approximate invariant, this approach
offers a scalable and efficient tool for nonlinear analysis.
It provides a compact, interpretable representation of complex
dynamics and opens a new pathway for nonlinear optimization and
control in next-generation accelerator systems, serving as both
a conceptual and computational bridge between practical machine
design and modern dynamical systems theory.

\section{Acknowledgments}

The authors would like to thank Taylor Nchako (Northwestern
University) for carefully reading this manuscript and for her
helpful comments.
The authors gratefully acknowledge Eric Stern, Robert Ainsworth
and Vladimir Nagaslaev (all from FermiLab) for providing the
lattice files for the IOTA, Main Injector, and Delivery Rings,
respectively.
S.N. work is supported by the U.S. Department of Energy,
Office of Science,
Office of Nuclear Physics under contract DE-AC05-06OR23177.
S.K. is grateful to his supervisor, Prof. Young-Kee Kim
(University of Chicago), for her valuable mentorship and continuous
support.

\vspace{0.5cm}
{\it Data Availability} ---
The data that support the findings of this study are available
from the corresponding author upon reasonable request.


\newpage

%

\end{document}